\documentclass[conference]{IEEEtran}
\IEEEoverridecommandlockouts
\usepackage{cite}
\usepackage{amsmath,amssymb,amsfonts}
\usepackage{algorithmic}
\usepackage{graphicx}
\usepackage{textcomp}
\usepackage{xcolor}
\usepackage{url,hyperref}

\usepackage {multirow}
\usepackage{makecell}
\newcommand{\eunwooedit}[1]{\textcolor{black}{#1}}
\newcommand{\minjaeedit}[1]{\textcolor{black}{#1}}
\newcommand{\minjaeeditt}[1]{\textcolor{black}{#1}}

\begin{document}

\title{LP-WaveNet: Linear Prediction-based \\ WaveNet Speech Synthesis \\
\thanks{Work partially performed when the first author was an intern at \minjaeedit{Microsoft Research Asia.}}
}

\makeatletter
\newcommand{\linebreakand}{%
\end{@IEEEauthorhalign}
\hfill\mbox{}\par
\mbox{}\hfill\begin{@IEEEauthorhalign}
}
\makeatother
\author{
\IEEEauthorblockN{Min-Jae Hwang}
\IEEEauthorblockA{
% \textit{Voice Team} \\
	\textit{Search Solution}\\
	Seongnam, South Korea \\
	min-jae.hwang@navercorp.com}
\and
\IEEEauthorblockN{Frank Soong}
\IEEEauthorblockA{
% \textit{Speech Group} \\
	\textit{Microsoft}\\
	Beijing, China \\
	frankkps@microsoft.com} 
\and
\IEEEauthorblockN{Eunwoo Song}
\IEEEauthorblockA{
% \textit{Voice Team} \\
	\textit{Naver Corporation}\\
	Seongnam, South Korea \\
	eunwoo.song@navercorp.com}
\linebreakand
\IEEEauthorblockN{Xi Wang}
\IEEEauthorblockA{
% \textit{Speech Group} \\
	\textit{Microsoft}\\
	Beijing, China \\
	xwang@microsoft.com} %\vspace*{-5mm}
\and
\IEEEauthorblockN{Hyeonjoo Kang}
\IEEEauthorblockA{
% \textit{Department of Electrical and Electronics} \\
	\textit{Yonsei University}\\
	Seoul, South Korea \\
	volleruhe@dsp.yonsei.ac.kr}% \vspace*{-5mm}
\and
\IEEEauthorblockN{Hong-Goo Kang}
\IEEEauthorblockA{
% \textit{Department of Electrical and Electronics} \\
	\textit{Yonsei University}\\
	Seoul, South Korea \\
	hgkang@yonsei.ac.kr} %\vspace*{-5mm}
}

\maketitle

\begin{abstract}

We propose a linear prediction (LP)-based waveform generation method via WaveNet vocoding framework.
A WaveNet-based neural vocoder has significantly improved the quality of parametric text-to-speech (TTS) systems.
However, it is challenging to effectively train the neural vocoder when the target database contains massive amount of acoustical information such as prosody, style or expressiveness.
As a solution, the approaches that only generate the vocal source component by a neural vocoder have been proposed.
However, they tend to generate synthetic noise because the vocal source component is independently handled without considering the entire speech production process; where it is inevitable to come up with a mismatch between vocal source and vocal tract filter.
To address this problem, we propose an LP-WaveNet vocoder, where the complicated interactions between vocal source and vocal tract components are jointly trained within a mixture density network-based WaveNet model.
The experimental results verify that the proposed system outperforms the conventional WaveNet vocoders both objectively and subjectively.
In particular, the proposed method achieves 4.47 MOS within the TTS framework.

\end{abstract}

\begin{IEEEkeywords}
Text-to-speech, speech synthesis, WaveNet vocoder  \vspace{-1mm}
\end{IEEEkeywords}

\section{Introduction}
%\IEEEPARstart{T}{he}
%The WaveNet vocoder \cite{Tamamori2017SpeakerDependentWV, hayashi2017investigation}, which uses acoustic features as a conditional input of WaveNet \cite{aaron2016WaveNet}, significantly improves the synthesis quality of conventional deep learning-based statistical parametric speech synthesis (SPSS) systems.
Waveform generation systems using WaveNet have significantly improved the synthesis quality of deep learning-based text-to-speech (TTS) systems \cite{oord2016wavenet, Tamamori2017SpeakerDependentWV, hayashi2017investigation, shen2018natural, adiga2018use}.
Because the WaveNet vocoder can generate speech samples in a single unified neural network, it does not require any hand-engineered signal processing pipeline.
Thus, it presents much higher synthetic quality than the traditional parametric vocoders \cite{Tamamori2017SpeakerDependentWV}.

To further improve the perceptual quality of the synthesized speech, more recent neural \textit{excitation} vocoders take advantages of the merits from both the linear prediction (LP) vocoder and the WaveNet structure \cite{lauri2018speaker, yang2018new, song2018neural, wang2018neural, Valin2018LPCNetIN}. 
In this framework, the formant-related spectral structure of the speech signal is decoupled by an LP analysis filter, and the WaveNet only estimates the distribution of its residual signal (i.e., excitation).
\minjaeeditt{Because the physical behavior of excitation signal is simpler than the speech signal, the training and generation processes become more efficient.}
%As a result, the training and generation processes become more efficient.}

% However, the synthesized speech is likely to be unnatural when the prediction errors \minjaeeditt{caused by excitation estimation} are propagated \minjaeedit2{through} the LP synthesis process.
However, the synthesized speech is likely to be unnatural when the prediction errors \minjaeeditt{in estimating the excitation} are propagated \minjaeeditt{through} the LP synthesis process.
\minjaeeditt{As} the effect of LP synthesis is not considered in the training process, the synthesis output is vulnerable to the variation of LP synthesis filter.

To alleviate this problem, we propose an \textbf{\textit{LP-WaveNet}}, which enables to jointly train the complicated interactions between the excitation and LP synthesis filter.
Based on the basic assumption that the past speech samples and the LP coefficients are given as conditional information, we figure out that the distributions of speech and excitation only lies on a constant difference.
Furthermore, if we model the speech distribution by using a mixture density network (MDN) \cite{Bishop94mixturedensity}, then the target speech distribution can be estimated by summing the mean parameters of predicted mixture \minjaeeditt{and} an \textit{LP approximation}, which is defined as the linear combination of past speech samples weighted by LP coefficients.
Note that the LP-WaveNet is easy to train because the WaveNet only needs to model the excitation \minjaeeditt{component}, and the complicated spectrum modeling \minjaeeditt{part} is embedded into the LP approximation.

In the objective and subjective evaluations, we verified the outperforming performance of \minjaeeditt{the} proposed LP-WaveNet in comparison to the conventional WaveNet-based neural vocoders.
Especially, the LP-WaveNet provided 4.47 mean opinion score result in the TTS framework.

\section{WaveNet-based speech synthesis systems}
\label{sec:background}
\subsection{$\mu$-law quantization-based WaveNet}
WaveNet is a convolutional neural network (CNN)-based auto-regressive generative model that predicts the joint probability distribution of speech samples $\mathbf{x}=\{x_{1}, x_{2}, ..., x_{N}\}$ as follows:
\begin{equation}
\label{eq:wavenet:prob}
p(\mathbf{x}| \mathbf{h})=\prod_{n}p(x_{n}|\mathbf{x}_{<n}, \mathbf{h}),
\end{equation}
where $x_{n}$, $\mathbf{x}_{<n}$, and $\mathbf{h}$ \eunwooedit{denote} the $n^{th}$ speech sample, its past speech samples, and the acoustic features, respectively. % are --> denote
By stacking the dilated causal convolution layers multiply, the WaveNet effectively extends its receptive field to the thousand of samples.

The firstly proposed WaveNet, a.k.a., $\mu$-law WaveNet \cite{oord2016wavenet}, defines the distribution of speech sample as a 256 categorical class of symbols obtained by an 8-bit $\mu$-law quantized speech samples.
To model the distribution of speech sample, the categorical distribution is computed by applying softmax operation to the output of WaveNet.
In the training phase, the weights of WaveNet is updated to minimize the cross-entropy loss.
In the generation phase, the speech sample is auto-regressively generated in sample-by-sample.

Since the $\mu$-law WaveNet can generate the speech signal in a single unified model, it provides significantly better synthetic sound than the conventional parametric \eunwooedit{vocoders.}
% which is constructed by heuristic and inaccurate signal processing pipeline.
% 공격적인 문장 뺌
However, it is not easy to train the network when the amount of database is larger and its acoustical informations such as prosody, style, or expressiveness are wider.
Moreover, the synthesized sound of WaveNet is often suffered from the background noise artifact as the target speech signal is too coarsely quantized.

\subsection{WaveNet-based excitation modeling}
\label{sec:excitation_modeling}
One effective solution is to model the \eunwooedit{excitation} signal instead of the speech signal. % vocal source  --> excitation
For instance, in the ExcitNet approach \cite{song2018neural},  an excitation signal is first obtained by an LP analysis filter, then its probabilistic behavior is trained by the WaveNet framework.

During the synthesis, the excitation signal is generated by the trained WaveNet, then it is passed through an LP synthesis filter to synthesize the speech signal as follows:
\begin{equation}
\label{eq:lp_syn}
\begin{aligned}
x_{n} & = e_{n} + \hat{x}_n, \\ 
\hat{x}_n & = \sum_{i=1}^{p}\alpha_{i}x_{n-i}, 
\end{aligned}
\end{equation}
where $e_n$, $\hat{x}_n$, $p$, and $\alpha_i$ denote the $n^{th}$ sample of excitation signal, the intermediate \textit{LP approximation} term, the order of LP analysis, and the $i^{th}$ LP coefficient, respectively.
Note that the LP coefficients are periodically updated to match with the extraction period of acoustic features.
For instance, if acoustic features are extracted at every 5-ms, then the LP coefficients are updated at every 5-ms to synchronize the feature update interval.

%Because the structure of excitation signal is simpler than that of speech signal, its training is much easier and the quality of finally synthesized speech is much higher, too.
Because the \eunwooedit{variation in the excitation signal is only constrained by vocal cord movement,} its training is much easier and the quality of finally synthesized speech is much higher, too.
However, the synthesized speech often contains unnatural artifacts because the excitation model is trained independently without considering the effect of LP synthesis filter; where it happens mismatch between the excitation signal and LP synthesis filter.
To address this limitation, we propose an LP-WaveNet, where both excitation signal and LP synthesis filter are jointly considered for training and synthesis.

\section{Linear prediction WaveNet vocoder}
\label{sec:lp_wavenet}
\subsection{Fundamental mathematics}
Before introducing the proposed LP-WaveNet, a probabilistic relationship between speech and excitation signals have to be clarified.
Note that at the moment of $n^{th}$ sample generation process in the WaveNet's synthesis stage, $\hat{x}_n$ shown in Eq.~\eqref{eq:lp_syn} can be treated as a given factor since both LP coefficients, $a_i$, and previously reconstructed samples, $\{x_{n-i}\}$ are already estimated.
Hence, we conclude that the difference between two random variables, $x_n$ and $e_n$, is only a known constant value term of $\hat{x}_n$.

Considering the shift property of second-order random variable, if we define the speech's distribution as a mixture of Gaussian (MoG), the relationship between mixture parameters of speech and excitation distributions can be lie on the only constant difference of mean parameters as follows:
\begin{equation}
\label{eq:mog}
p(x_n | \mathbf{x}_{<n}, \mathbf{h}) = \sum_{i=1}^{N} \frac{w_{n,i}}{\sqrt{2\pi}s_{n,i}} \text{exp} \left[ -\frac{(x_n-\mu_{n,i})^2}{2s_{n,i}^2} \right],
\end{equation} \vspace{-2mm}
\begin{gather}
\begin{aligned}
\label{eq:relation3}
w_{n,i}^x =& w_{n,i}^e, \\	
\mu_{n,i}^x =& \mu_{n,i}^e   + p_n, \\
s_{n,i}^x =& s_{n,i}^e, 
\end{aligned}	
\end{gather}
where $N$ and $i$ denote the number and index of \eunwooedit{mixture, respectively}; $w$ denotes the weights of mixture component; $\mathcal{N}(\mu, s)$ imply the Gaussian distribution having mean of $\mu$ and standard deviation of $s$; the superscripts $e$ and $x$ denote the excitation and the speech, respectively.
Based on this observation, we propose an LP-WaveNet vocoder, where the LP synthesis process is structurally reflected to the WaveNet's training and inference processes.

\subsection{Network architecture}
The detailed architecture of LP-WaveNet is illustrated in Fig.~\ref{fig:lp_wavenet}.
In the proposed system, the distribution of speech sample is defined as a MoG distribution by following Eq.~\eqref{eq:mog}, and the LP-WaveNet is trained to generate the MoG parameters, $[\boldsymbol{w}_n, \boldsymbol{\mu}_n, \boldsymbol{s}_n]$ conditioned by the input acoustic features.

\begin{figure}[t!]
	\centering
	\includegraphics[width=0.9\linewidth]{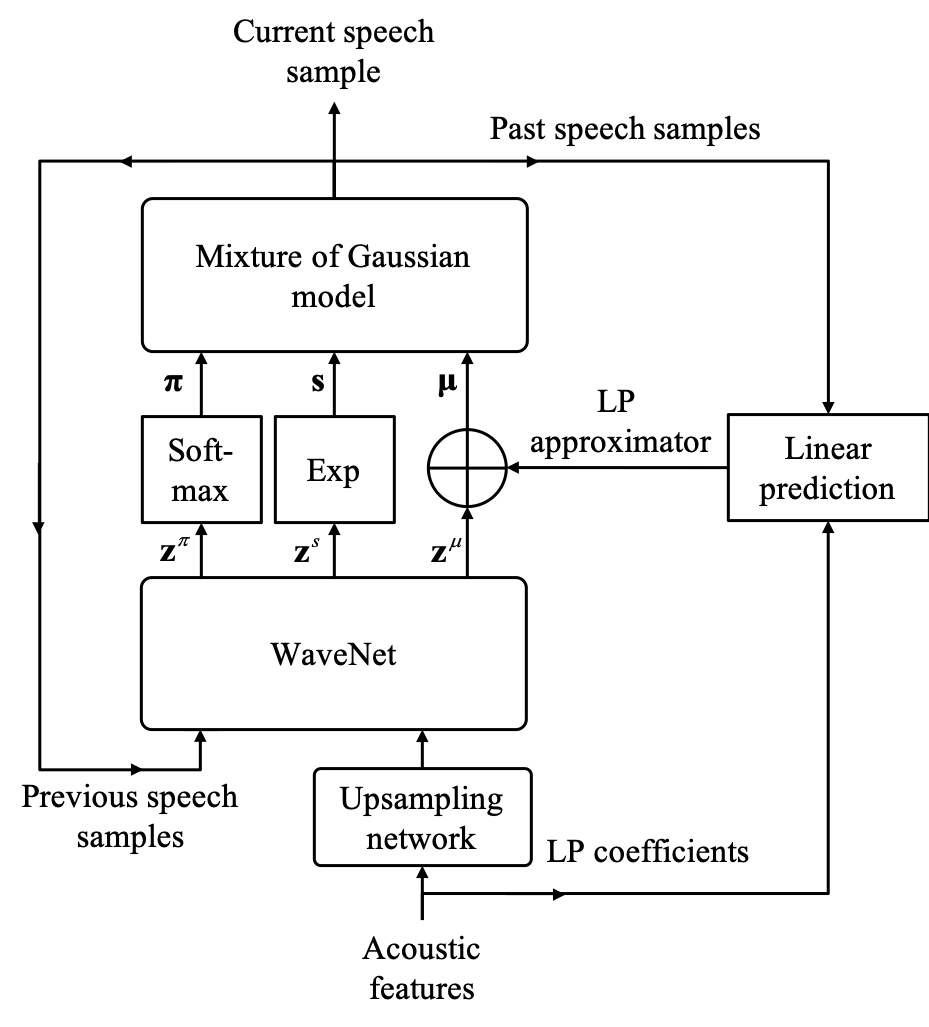}
	\caption{
		Block diagram of the LP-WaveNet vocoder.
	}
	%		\vspace{-1mm}
	\label{fig:lp_wavenet}
\end{figure}

In detail, the acoustic features pass through two \minjaeeditt{1-dimensional convolution layers having kernel size of 3} for explicitly imposing the contextual information of feature trajectory.
% Firstly, the acoustic features are upsampled to match its time resolution to that of speech signal.
% In detail, the acoustic features pass through two \minjaeeditt{1-dimensional convolution layers with kernel size of 3}, whose filter is extended along with the time axis for explicitly imposing the contextual information of feature trajectory.
Then, the residual connection with respect to the input acoustic feature is applied to make the network more focus on the current frame information.
Finally, the transposed convolution is applied to upsample the temporal resolution of this features into that of speech signal.

To generate the speech samples, the mixture parameters, i.e., mixture gain, mean and log-standard deviation, of excitation signal are first predicted by WaveNet as follows:
\begin{gather}
[\mathbf{z}^w_n, \mathbf{z}^\mu_n, \mathbf{z}^s_n] = WaveNet(\mathbf{x}_{<n}, \mathbf{h}_n)
\end{gather}
Then, the LP approximation term, $\hat{x}_n$, is computed by following Eq.~\eqref{eq:lp_syn} \eunwooedit{to generate the MoG parameters of speech sample as follows:}
%Then, the LP approximation term, $\hat{x}_n$, is computed by following Eq.~\eqref{eq:lp_syn}.
%Finally, the MoG parameters of speech sample are generated as follows:
\begin{gather}
\boldsymbol{\pi}_n = \text{softmax} (\mathbf{z}^\pi_n)  \nonumber \\
\boldsymbol{\mu}_n= \mathbf{z}^\mu_n + \hat{x}_{n}, \label{eq:lp_wavenet_mixture} \\ 
\boldsymbol{s}_n = \text{exp}(\mathbf{z}^{s}_n).   \nonumber
\end{gather}
Finally, the likelihood of speech sample $p(x_n | \mathbf{x}_{<n})$ is computed by following Eq.~\eqref{eq:mog}.

To train the network, the negative log-likelihood (NLL) of speech signal, $\mathcal{L}$, is computed from the MoG distribution defined at Eq.~\eqref{eq:mog} as follows:
\begin{gather}
\label{nll_mog}
\begin{aligned}
\mathcal{L} & = -\text{log} p(x_n | \mathbf{x}_{<n}).
\end{aligned}
\end{gather}
Then, the weights are optimized to minimize NLL loss.

Because the complicated spectral modeling is now embedded in the mean parameters as depicted in Eq.~\eqref{eq:lp_wavenet_mixture}, the LP-WaveNet only needs to train an information of excitation signal, which is relatively easy to train.
Moreover, because the ultimate training target of LP-WaveNet is speech signal, it is also free from the mismatch problem mentioned in Section~\ref{sec:excitation_modeling} 
As a result, the LP-WaveNet is able to model the both excitation generation and LP synthesis filter processes jointly in a WaveNet structure.

\subsection{Backpropagation method}
Note that the constant summation guarantees the linearity.
Thus, the weights of WaveNet can be successfully trained by a standard back-propagation process.
In detail, the partial derivative of NLL loss $\mathcal{L}$ with respect to $\boldsymbol{z}^{\mu}_n$ can be represented as follows:
\begin{equation}
\label{eq:nll_to_z}
\frac{\partial \mathcal{L}}{\partial z_{i}^{\mu}} = \frac{\partial \mathcal{L}}{\partial \mu_{j}} \cdot \frac{\partial \mu_{j}}{\partial z_{i}^{\mu} }.
\end{equation}
The time index $n$ is omitted for the readability.
Note that $\hat{\mu}_j$ and $z_i^{\mu}$ are coupled by the constant difference and their off-diagonal components ($i \neq j$) are decorrelated.
Thus, their partial derivative becomes:
\begin{equation}
\label{eq:mu_to_z}
\frac{\partial \mu_j}{\partial z_i^{\mu} } = \delta(i-j),
\end{equation}
where $\delta(\cdot)$ is the Dirac-delta function.
Thus, Eq.~\eqref{eq:nll_to_z} is simplified as follows:
\begin{equation}
\label{eq:nll_to_mu}
\frac{\partial C}{\partial z_i^{\mu}} = \frac{\partial C}{\partial \mu_i}.
\end{equation}
This implies that the back-propagation of LP-WaveNet can be easily implemented by assigning the generated mean components to the back-propagation rule of MDN, which is already defined \cite{Bishop94mixturedensity}.

\section{Effective training and generation methods}

%\subsection{Training noise injection}
%One of the WaveNet's problem during its training is that the WaveNet is easy to be over-fitted to the silence region.
%In the training phase, the WaveNet is forced to learn the unique solution at the silence region.
%Thus, the WaveNet is over-fitted on this unique solution as the amount of silence region increases.
%In the synthesis stage, the generated speech that contains a small value of error even in the silence region, is auto-regressively fed to the next input.
%Because the WaveNet is over-fitted on the silence region, this small value of error term is crucial an results in error propagation and clipping noise.
%
%To solve this problem, we add an imperceptibly negligible artificial noise to the speech signal.
%Considering the 16-bit linearly quantized audio signal, the training noise is designed to have zero-mean and standard deviation of $2/2^{16}$.

\subsection{Waveform generation via conditional distribution sharpening}
During waveform generation, a random sampling that follows the probability distribution of waveform is commonly used.
However, its synthetic sound is noisy due to the stochastic sampling process.
%When the WaveNet's output layer is a softmax layer, this noise can be controlled by adjusting the sharpness of waveform distribution \cite{zeyu2018FFTNet, xin2018comparison}.
%However, there's no prior studies that using this solution on the MDN-WaveNet.
In this study, we control the noisiness by adjusting the sharpness of waveform distribution by reducing the scale parameters generated by the WaveNet.
Because the buzziness and the hiss of synthetic speech are sensitive to the sharpness of distribution, the scale parameters have to be carefully adjusted.
After several trials, we concluded that reducing the scale by factor of 0.85 at only voiced region presents the best performance.

\subsection{Upper bound limitation on the generated log-scale parameters}
During the waveform generation process, we figured out that the generated waveform can be often unstable when the generated log-scale parameters are too high.
%To prevent this problem, the generated log-scale parameters were bounded up to --4 log scale value.
%In the waveform generation process of WaveNet, we figured out that the waveform generation process often became unstable in the unvoiced region \cite{yi2018collapsed}.
%In detail, the unexpectedly exploding noise artifact is sometimes happened when the scale parameters became abnormally large at the unvoiced region.
This problem could be prevented by clipping the upper bound of scale parameter value.
If the clipping was set too low, then the unvoiced region was not sufficiently modeled, resulting in a dry synthetic sound though the waveform could be stably generated.
If the clipping was set too high, then the possibility of waveform explosion became higher, but the synthetic sound became more lively than the lower clipping value case.
Based on experiments, we limited the scale parameter to $-4.0$ natural logarithm.

\section{Experiments}
\label{sec:experiment}
\subsection{Speech database and features}
In the experiments, phonetically and prosodically riching speech corpus recorded by a professional Korean female speaker was used for the experiments.
The speech signals were sampled at 24-kHz with 16-bits quantization.
The randomly selected 4,976 utterances \eunwooedit{(9.9 hours)} were used for training, 280 utterances were used for validation, and another 140 utterances were used for test, respectively.
The acoustic features were obtained by the ITFTE vocoder \cite{song2017effective} at every 5-ms interval; 40-dimensional \eunwooedit{line spectral frequencies (LSFs), logarithmic fundamental frequency (F0), logarithmic energy, voicing flag, 32-dimensional slowly evolving waveform, and 4-dimensional rapidly evolving waveform, all of which composed a total 79-dimensional feature vector}.
%\vspace{-3mm}		
\subsection{WaveNet vocoders}
Total three WaveNet vocoding systems were tested.
\begin{itemize}
	\item WN$_{S}$: $\mu$-law WaveNet vocoder that directly models the speech signal 
	\cite{Tamamori2017SpeakerDependentWV}.
	\item WN$_{E}$: ExcitNet vocoder that models the excitation signal with explicit LP synthesis filter \cite{song2018neural}.
	\item WN$_{LP}$: Proposed LP-WaveNet vocoder.
\end{itemize}

For a fair comparison with similar computing resource, the same WaveNet architecture was used to all systems.
Firstly, the dilations were set to [$2^0, 2^1, ..., 2^9$] and repeated three times, resulting in 30 layers of residual blocks and 3,071 samples of the receptive field.
In the residual blocks and the post-processing module, the 128 channels of convolution layers were used.
The number of mixture was set to 10, resulting in 30 channels of output layer.
For the LP-WaveNet, the single Gaussian distribution was assumed, and the weight normalization technique, which normalizes the weight vectors to have unit-length, is applied to stabilize a training process of LP-WaveNet \cite{salimans2016weight}.
Moreover, the scale parameter was clipped by the lower bound of $-10.0$ natural logarithm when calculating a negative log-likelihood (NLL) loss to stabilize the training of mixture density network (MDN) \cite{ping2018clarinet}.
The weights were firstly initialized by the \textit{Xavier} initializer \cite{xavier2010init}, and then trained using an \textit{Adam} optimizer \cite{diederik2014adam}.
The learning rate was set to $10^{-4}$.
The mini-batch size was 20,000 samples with 8GPUs, resulting in 160,000 samples per mini-batch.
The networks were trained in 600,000 iterations.

%	\vspace{-3mm}
\subsection{TTS acoustic model}
\label{sec:acoustic_model}
%To test the performance of WaveNet vocoders in an SPSS system, the hybrid network of feed-forward (FF) network and long-short term memory (LSTM) network was used as an acoustic model.
%In detail, the three feed-forward (FF) layers with 1,024 dimensional hidden layer were used at the input side, and the one LSTM layer with 512 memory cells was used at the output side.
To \eunwooedit{evaluate} the performance of WaveNet vocoders in the TTS system, \eunwooedit{we implemented a simple acoustic model by using multiple feed-forward (FF) and long-short term memory (LSTM) layers.}
In detail, \eunwooedit{the network consisted of three FF layers with 1,024 units and one LSTM layer with 512 memory cells}.
The ReLu activation and linear functions were used at the hidden and output layers, respectively.

%As an input vector, total 356-dimensional linguistic features including 330 binary features of categorical linguistic contexts and 26 numerical features of numerical linguistic contexts.
\eunwooedit{The input vector was composed of 356-dimensional linguistic features including 330 binary features of categorical linguistic contexts and 26 numerical features of numerical linguistic contexts.}
\eunwooedit{The corresponding output vector consisted of all the acoustic parameters together with their time dynamics \cite{furui1986speaker}.
Before training, both input and output features were normalized to have zero mean and unit variance.
The weights were trained using a backpropagation through time algorithm with Adam optimization \cite{williams1990efficient}.}

\eunwooedit{In the synthesis step, the means of all acoustic features were predicted by the acoustic model first, then a speech parameter generation algorithm was applied with the pre-computed global variances \cite{tokuda2000speech}. 
To enhance spectral clarity, an LSF-sharpening filter was also applied to the spectral parameters \cite{song2017effective}. 
Finally, the generated acoustic features were used to compose the input features of the WaveNet vocoders.
}
\subsection{Objective and subjective evaluation results}
\begin{table}[t!]
	\centering
	\caption{Objective evaluation results of the various WaveNet vocoders with analysis and synthesis (A/S) and \eunwooedit{parametric TTS} systems.
		The system with highest performance is represented in bold typeface.}
	\label{table:objective_result}	
	%				\small
	%	\renewcommand{\arraystretch}{1.0 5}
	%		\setlength\tabcolsep{4.5pt} % default value: 6pt
	\begin{tabular}{c||c||c|c|c|c}
		\Xhline{2\arrayrulewidth}
		& \multirow{2}{*}{System}    & VUV           & F0 RMSE       & LSD           & F-LSD         \\ 
		& 				& (\%)          & (Hz)          & (dB)          & (dB)          \\ \hline\hline
		\multirow{3}{*}{A/S} 		&	WN$_{S}$   &   4.09      & 3.76          & 2.01          & 9.90           \\ 
		& WN$_{E}$   & 3.77          & 3.17         & 2.32 & 8.80          \\ 
		&WN$_{LP}$  & \textbf{2.28} & \textbf{2.70} & \textbf{1.67}          & \textbf{8.47}  \\ \hline
		\multirow{3}{*}{TTS} 		&	WN$_{S}$   & 5.06    & 13.67         & 4.45          & 12.81         \\ 
		& WN$_{E}$   & 4.84          & 13.61         & 4.43 & \textbf{12.30}         \\ 
		&WN$_{LP}$  & \textbf{4.12}     & \textbf{13.54}& \textbf{4.41}          & 12.37	\\ \Xhline{2\arrayrulewidth}
	\end{tabular}
%\vspace{-3mm}
	\label{sec:objective_results}
\end{table}

In the objective test, distortions in acoustic features extracted by the original speech and synthesized speech were evaluated.
Firstly, the analysis and synthesis (A/S) system, which synthesizes the speech with the ground truth acoustic features was tested to evaluate the vocoder's performance itself.
Then, the TTS system, which uses the acoustic features predicted by the LSTM-based acoustic condition model was tested in a real application scenario.

The metrics for the distortion measuring were the error rate of voicing flag (VUV) in \%, the root mean square error (RMSE) for F0 in Hz, the log-spectral distance (LSD) for LSFs in dB, and the LSD for speech magnitude response in frequency domain (F-LSD) in dB.
All the features needed for the metrics were extracted with 35-ms window at every 5-ms interval, then all the measures were averaged.
The F0 RMSE and F-LSD were measured in only voiced region.
To estimate the F-LSD, by computing phase mismatch, we compensated a lag to have maximum correlation between two speech frames within a 5-ms sample shift interval.

The objective evaluation of A/S and TTS results are summarized in Table~\ref{table:objective_result}.
The experimental results verify that 
(1) In all matrices, the proposed WN$_{LP}$ showed significantly better performance than the conventional WN$_S$ and WN$_{E}$ when the acoustic features are ground truth.
(2) All the performances are degraded in the TTS system as the prediction error of acoustic features. However, the performance of LP-WaveNet is still signifcantly better than the other systems.

\begin{table}[t!]
	\centering
%	\normalsize
	\caption{Subjective mean opinion score (MOS) test result with a 95\% confidence interval for various speech synthesis systems. The system with highest score is represented in bold typeface. The MOS result of recorded speech was 4.75.} %$\pm$0.05
	\label{table:subjective_mos}
	\begin{tabular}{c||c|c|c|c}
		\Xhline{2\arrayrulewidth}
		& ITFTE & WN$_{S}$ & WN$_{E}$ & WN$_{LP}$     \\ \hline\hline
		A/S & 2.85$\pm$0.20 & 3.40$\pm$0.19     & 4.11$\pm$0.16     & \textbf{4.58$\pm$0.12} \\ \hline
		TTS & 2.32$\pm$0.06 & 3.57$\pm$0.11    & 4.04$\pm$0.16     & \textbf{4.47$\pm$0.09} \\ \Xhline{2\arrayrulewidth}
	\end{tabular}
%\vspace{-3mm}
\end{table}

\begin{table}[t!]
	\centering
	\caption{
		Subjective preference test results (\%) between various WaveNet vocoding systems.
		The systems that achieved significantly better preference at the $p < 0.01$ level are in bold typeface.
	}
	\label{table:subjective_abx}
	\begin{tabular}{c|c||ccc|c||c}
		\Xhline{2\arrayrulewidth}
%		\multirow{2}{*}{Test index} & \multicolumn{3}{c|}{A/S}        & \multicolumn{3}{c|}{SPSS}       & \multirow{2}{*}{Neutral} & \multirow{2}{*}{$p$-value} \\ \cline{2-7}
		Index   			& System  & WN$_{S}$ & WN$_{E}$ & WN$_{LP}$ &     Neutral    &    $p$-value         \\ \hline\hline
		Test 1             &  \multirow{3}{*}{A/S}       & \textbf{9.4}      & \textbf{71.3}     & --        & \textbf{19.3}                     & $\mathbf{<10^{-21}}$                 \\ 
		Test 2             &         & \textbf{2.6}      & --       & \textbf{82.7 }     & \textbf{14.7}                     & $\mathbf{<10^{-45}}$                     \\ 
		Test 3             &        & --       & \textbf{12.0}     & \textbf{52.0}      & \textbf{36.0}                     & $\mathbf{<10^{-10}}$                     \\ \hline
		Test 4             &  \multirow{3}{*}{TTS}      & \textbf{8.0}      & \textbf{57.3}    & --        & \textbf{34.7}                      & $\mathbf{<10^{-16}}$                     \\ 
		Test 5              &       &  \textbf{1.4}      & --       & \textbf{79.3}     & \textbf{19.3}                      & $\mathbf{<10^{-46}}$                     \\ 
		Test 6              &       & --       & \textbf{12.0}      & \textbf{33.3}      & \textbf{54.7}                    & $\mathbf{<10^{-4}}$                 \\ 		
		\Xhline{2\arrayrulewidth}
	\end{tabular}
%\vspace{-3mm}
\end{table}

To evaluate the perceptual quality of the proposed system, the mean opinion score (MOS) listening test A-B preference test were performed\footnote{\minjaeedit{Generated audio samples are available at the following URL:	\\ \url{https://min-jae.github.io/eusipco2020/}}}.
Total 11 native Korean listeners were asked to score the randomly selected 15 synthesized utterances from the test set using a following possible 5-point MOS responses: 1 = Bad, 2 = Poor, 3 = Fair, 4 = Good, 5 = Excellent.
In addition to the WaveNet vocoding systems, the ITFTE-based vocoding system \cite{song2017effective}, i.e., ITFTE, having the same acoustic model with the WaveNet vocoding system was also included as a reference system.

The MOS test results are summarized in Table~\ref{table:subjective_mos}.
In the A/S system, all of WaveNet vocoders showed better quality than the parametric ITFTE vocoder.
Specifically, the proposed WN$_{LP}$ showed the best quality among the WaveNet vocoders with the only 0.17 lower MOS score than the recorded speech.
Even though the MOS result of WN$_E$ was higher than 4.0, its quality was significantly worse that the proposed WN$_{LP}$.
In the TTS system, all systems presented worse synthetic quality than the A/S system due to the prediction error of acoustic features.
However, their relative tendency was same with the results of A/S system.
Even though the prediction error of acoustic features, the proposed WN$_{LP}$ showed very high quality of synthesized speech with 4.47 MOS.
% whereas the quality of WN$_{E}$ was significantly degraded below 4.0 MOS.
% This implies that the proposed WN$_{LP}$ has stronger robustness than the conventional WN$_E$.

The setup for the A-B preference test was the same as that for the preference test, except the listeners were asked to rate the randomly selected 15 synthesized utterances from the test set by a quality preference. 
The preference results shown in Table~\ref{table:subjective_abx} verified that the perceptual quality of the proposed WN$_{LP}$ was significantly better than the conventional WN$_E$ and WN$_S$ in both of A/S and TTS systems (Test 2, 3, 5, and 6).
Also, the WN$_E$ verified that its performance was better than the plain WN$_S$ (Test 1 and 4).

%\vspace{-3mm}
\section{Conclusion}
\label{sec:conclusion}
In this paper, we proposed an LP-WaveNet vocoder.
By utilizing the causality of WaveNet and the linearity of LP synthesis filtering process, we structurally merged the LP synthesis filter into the WaveNet framework.
The experimental results verified that the proposed system outperformed the conventional WaveNet systems both objectively and subjectively.
\minjaeedit{Future works include to extend the idea of LP-WaveNet to the non-autoregressive waveform models for achieving real-time waveform generation.}
\section{Acknowledgements}
%\vspace{-2mm}
The work was supported by Clova Voice, NAVER Corp., Seongnam, Korea.

%\vspace{3mm}
%\vfill
%\pagebreak
\bibliographystyle{IEEEtran}
%	\fontsize{9.5}{11}\selectfont
%\fontsize{10}{11.5}\selectfont
\bibliography{mybib_mj}	
\end{document}